\documentclass{JHEP3}

\usepackage{amsmath,amssymb}
\usepackage[dvips]{graphics}
\def\beq{\begin{equation}}
\def\eeq{\end{equation}}
\def\bea{\begin{eqnarray}}
\def\eea{\end{eqnarray}}
\title{Partition functions and double-trace deformations in AdS/CFT}
\author{Danilo E. D\'{\i}az and Harald Dorn
\\ Humboldt-Universit\"at zu Berlin, Institut f\"ur Physik
\\Newtonstr.15, D-12489 Berlin
\\E-mail: \email{ddiaz,dorn@physik.hu-berlin.de}}

\abstract{ We study the effect of a relevant double-trace deformation
on the partition function (and conformal anomaly) of a $CFT_d$ at
large $N$ and its dual picture in $AdS_{d+1}$. Three complementary
previous results are brought into {\em full} agreement with each
other: bulk~\cite{GM03} and boundary~\cite{GK03} computations, as
well as their formal identity~\cite{HR06}. We show the {\em exact}
equality between the dimensionally regularized partition functions
or, equivalently, fluctuational determinants involved. A series of
results then follows: {\it(i)} equality between the renormalized
partition functions for all $d$; {\it(ii)} for {\it all} even $d$,
correction to the conformal anomaly; {\it(iii)} for even $d$, the
mapping entails a mixing of UV and IR effects on the same side
(bulk) of the duality, with no precedent in the leading order
computations; and finally, {\it(iv)} a subtle relation between
overall coefficients, volume renormalization  and IR-UV connection.
All in all, we get a clean test of the AdS/CFT  correspondence
beyond the classical SUGRA approximation in the bulk and at
subleading $O(1)$ order in the large-$N$ expansion on the boundary.}
\preprint{HU-EP-07/04}

\begin{document}
\section{Introduction}

Maldacena's conjecture and its calculational
prescription~\cite{Malda} entail the equality between the partition
function of String/M-theory (with prescribed boundary conditions) in
the product space $AdS_{d+1}${\tiny $\times$} $X$, where $X$
is certain compact manifold,  and the generating functional of the
boundary $CFT_d$. It has been fairly well tested at the level of
classical SUGRA in the bulk and at the corresponding leading order
at large $N$ on the boundary.

One of the most remarkable tests is the mapping of the conformal
anomaly~\cite{HS98}. Since the rank $N$ of the group measures the
size of the geometry in Planck units, quantum corrections correspond
to subleading terms in the large $N$ limit. Corrections of order
$O(N)$ have also been obtained~\cite{ON}, but they rather correspond
to tree-level corrections after inclusion of open or unoriented
closed strings. Truly  quantum corrections face the notorious
difficulty of RR-backgrounds and only few examples,  besides
semiclassical limits of the correspondence, have circumvented it
and corroborated the conjecture at this nontrivial
level~\cite{Quant}. These results rely on whole towers of KK-states
and SUSY. The regimes in which the bulk and boundary computations
can be done do not overlap and some sort of non-renormalization must
be invoked.

In this note we deal with a universal AdS/CFT result, not relying on
SUSY or any other detail encoded in the compact space $X$,
concerning an $O(1)$ correction to the conformal anomaly under a
flow produced by a double-trace deformation. This was first computed
in the bulk of AdS~\cite{GM03} and confirmed shortly after by a
field theoretic computation on the dual boundary theory~\cite{GK03}
(see also~\cite{Gub04}).

Let us roughly recapitulate the sequence of developments leading to
this remarkable success. It starts with a scalar field $\phi$ with
``tachyonic'' mass in the window $-\frac{d^2}{4}\leq
m^2<-\frac{d^2}{4}+1$ where two AdS-invariant quantizations are
known to exist~\cite{BF82}. The conformal dimensions of the dual CFT
operators, given by the two roots $\Delta_+$ and $\Delta_-$ of the
AdS/CFT relation $m^2=\Delta(\Delta-d)$: \beq \Delta
_{\pm}~=~\frac{d}{2}\pm\nu~,~~~~~\nu~=~\sqrt{\frac{d^2}{4}+m^2}\label{delta}
\eeq are then ($0\leq \nu <1$) both above the unitarity bound. The
modern AdS/CFT interpretation~\cite{KW99} assigns the same bulk
theory to two different CFTs at the boundary, whose generating
functionals are related to each other by Legendre transformation at
leading large N. The only difference is the interchange of the roles
of boundary operator/source associated to the asymptotic behavior of
the bulk scalar field near the conformal boundary.
\\
The whole picture fits into the generalized AdS/CFT prescription to
incorporate boundary multi-trace operators~\cite{Multi}. The two
CFTs are then the end points of a RG flow triggered by the relevant
perturbation $f\,O_{\alpha}^2$ of the $\alpha-$CFT, where the
operator $O_{\alpha}$ has dimension $\Delta_-$ (so that
$\Delta_-+\Delta_+=d,\;\Delta_-\leq \Delta_+ \Rightarrow
2\Delta_-\leq d$). The $\alpha-$theory flows into the $\beta-$theory
which now has an operator $O_{\beta}$ with dimension
$\Delta_+=d-\Delta_-$, conjugate to $\Delta_-$. The rest of the
operators remains untouched at leading large $N$, which suggests
that the metric and the rest of the fields involved should retain
their background values, only the dual bulk scalar changes its
asymptotics~\footnote{The simplest realization of this behavior
being the $O(N)$ vector model in $2<d<4$, see
e.g.~\cite{KP02,DeD06}.}.

The crucial observation in~\cite{GM03} is the following: since the
only change in the bulk is in the asymptotics of the scalar field,
the effect on the partition function cannot be seen at the classical
gravity level in the bulk, i.e. at leading large $N$, since the
background solution has $\phi=0$; but the quantum fluctuations
around this solution, given by the functional determinant of the
kinetic term (inverse propagator), are certainly sensitive to the
asymptotics since there are two different propagators $G_{\Delta}$
corresponding to the two different AdS-invariant quantizations. The
partition function including the one-loop correction is \beq
Z^{\pm}_{grav}=Z^{class}_{grav}\;\cdot \left[ \mbox{det}_{\pm}
(-\Box + m^2) \right]^{-\frac{1}{2}},\eeq where $Z^{class}_{grav}$
refers to the usual saddle point approximation. Notice that the
functional determinant is independent of $N$, this makes the scalar
one-loop quantum correction an $O(1)$ effect. The 1-loop computation
turns out to be very simple for even dimension $d$ and is given by a
polynomial in $\Delta$. No infinities besides the IR one, related to
the volume of $AdS$, show up in the relative change $Z^{+}_{grav}/
Z^{-}_{grav}$, since the UV-divergences can be controlled exactly in
the same way for both propagators. From this correction to the
classical gravitational action one can read off an  $O(1)$
contribution to the {\em holographic conformal anomaly}~\cite{HS98}.

The question whether this $O(1)$ correction to the anomaly could be
recovered from a pure $CFT_d$ calculation was answered shortly after
in the affirmative~\cite{GK03}. Using the Hubbard-Stratonovich
transformation (or auxiliary field trick) and large $N$
factorization of correlators, the Legendre transformation relation
at leading large $N$ is shown. An extra $O(1)$ contribution, the
fluctuation determinant of the auxiliary field, is also obtained.
Turning the sources to zero, the result for the CFT partition
function can be written as \beq Z_{\beta}=Z_{\alpha}\;\cdot
\left[\mbox{det} (\Xi) \right]^{-\frac{1}{2}}, \eeq where the kernel
$\Xi\equiv\mathbb{I}+f \,\langle O_{\alpha}O_{\alpha}\rangle $ in
position space in $\mathbb{R}^d$ is given by $\delta^d(x,x')+
f\frac{C_{\Delta_-}}{\mid x-x'\mid^{2\Delta_-}}$. The $\beta-$CFT is
reached in the limit $f\rightarrow \infty$ .
\\
From the CFT point of view, the conformal invariance of  this
functional determinant has then to be probed. Putting the theory on
the sphere $\mathbb{S}^d$ and expanding in spherical harmonics,
using Stirling formula for large  orbital  quantum number $l$ and
zeta-function regularization, the coefficient of the log-divergent
term is isolated. It happily coincides (for the explored cases
$d=2,4,6,8$) with the $AdS_{d+1}$ prediction for the anomaly.

Despite the successful agreement, there are several issues in this
derivation that ought to be further examined. No track is kept on
the overall coefficient in the CFT computation, in contrast to the
mapping at leading order~\cite{HS98} that matches the overall
coefficient as well. For odd dimension $d$, the CFT determinant has
no anomaly, whereas there is a nonzero AdS result that could be some
finite term in field theory not computed so far \cite{GK03}. From a
computational point of view the results are quite different. The AdS
answer  is a polynomial for generic even dimension, whereas for odd
d only numerical results are reported. The CFT answer, on the other
hand, is obtained for few values of the dimension $d$, a proof
for generic $d$ is lacking. Yet, the very same $O(1)$ nature of the
correction on both sides of the correspondence calls for a full
equivalence between the relative change in the partition functions,
and not only just the conformal anomaly. This poses a new challenge
since in the above derivation there are several (divergent) terms
that were disregarded, for they do not contribute to the anomaly.

As we have seen, it all boils down to computing functional
determinants. In a more recent work~\cite{HR06}, a ``kinematical''
understanding of the agreement between the bulk and boundary
computations was achieved based on the equality between the
determinants. The key is to explicitly separate the transverse
coordinates in AdS, expand the bulk determinant in this basis
inserting the eigenvalues of the transverse Laplacian weighted with
their degeneracies. In this way, one gets a weighted sum/integral of
effective radial (one-dimensional) determinants which are then
evaluated via a suitable generalization of the Gel'fand-Yaglom
formula. The outcome turns out to coincide with the expansion of the
auxiliary field fluctuation determinant of~\cite{GK03}. However,
this procedure is known to be rather formal (see, e.g.,~\cite{DK06}
and references therein) and the result to be certainly divergent. No
further progress is done on either side of this {\em formal}
equality and the issue of reproducing the full bulk result from a
field theoretic computation at the boundary remains open.

We will show that all above open questions can be thoroughly
clarified or bypassed if one uses dimensional regularization to
control all the divergences. Both IR and UV divergencies are now on
equal footing, which is precisely the essence of the IR-UV
connection~\cite{WS98}: the key to the holographic bound is that an
IR regulator for the boundary area becomes an UV regulator in the
dual CFT. The bulk effective potential times the infinite AdS
volume, i.e. the effective action, and the boundary sum, using
Gau{\ss}'s ``proper-time representation'' for the digamma function
to perform  it, are shown to coincide in dimensional regularization.

The paper is organized as follows: we start with the bulk partition
function and compute the regularized effective action. Here one
needs to compute separately the volume and the effective potential.
Then we move to the boundary to compute the change induced by the
double-trace deformation. Having established the equivalence for
dimensionally regularized quantities we go back to the physical
dimensions and extract the relevant results for the renormalized
partition functions and the conformal anomaly. Before the final
conclusions, the paper still has a section with some further
technical remarks. Some useful relations and formulas are collected
in two appendices.

\section{The bulk computation: one-loop effective action}

Let us start with the Euclidean action for gravity and a scalar
field \beq S^{class}_{d+1}=\frac{-1}{2\kappa^2} \int d{\it {\it
vol}}_{d+1}\;[R-\Lambda] \quad+\quad \int d{\it {\it
vol}}_{d+1}\;[\frac{1}{2}(\nabla\phi)^2 + \frac{1}{2}m^2 \phi ^2].
\eeq For negative $\Lambda$ the Euclidean version of $AdS_{d+1}$,
i.e. the Lobachevsky space $\mathbb{H}^{d+1}$, is a classical
solution. There are, of course, additional terms like the
Gibbons-Hawking surface term\footnote{ Additional covariant local
boundary counterterms that render the on-shell classical bulk action
finite are obtained by the `holographic renormalization' procedure,
see e.g.~\cite{hr}.}  and contributions from other fields, but they
will play no role in what follows, nor will the details of the
leading large N duality. This is an indication of universality of
the results.

We are interested in the quantum one-loop correction from the scalar
field with the $\Delta_+$  or $\Delta_-$   asymptotic behavior

\beq S^{\pm}_{d+1}= \frac{1}{2}~\mbox{log\;det}_{\pm}(-\Box + m^2)=
\frac{1}{2}~\mbox{tr}_{\pm }\;\mbox{log}(-\Box + m^2)~.\eeq It will
prove simpler to consider instead the quantities \beq
\frac{\partial}{\partial
m^2}\;S^{\pm}_{d+1}=\frac{1}{2}~\mbox{tr}_{\pm}\;\frac{1}{-\Box +
m^2}~.\eeq  This rather symbolic manipulation, casted into a
concrete form, reads \beq \frac{\partial}{\partial
m^2}\;S^{\pm}_{d+1}=\frac{1}{2}\int d{\it
vol}_{\mathbb{H}^{d+1}}\;G_{\Delta_{\pm}}(z,z).\eeq There are two
kinds of divergencies here, one is the infinite volume of the
hyperbolic space (IR) and the other is the short distance
singularity of the propagator (UV). The latter is conventionally
controlled by taking the difference of the $\pm$-versions; this
produces a finite result and was the crucial observation in
\cite{GM03}. Then one gets for the difference of the one-loop
corrections for the $\Delta _{\pm}$ asymptotics \beq
\frac{\partial}{\partial m^2}\;\left (S^+_{d+1}-S^-_{d+1}\right )
=\frac{1}{2}\int d{\it vol}_{\mathbb{H}^{d+1}}\;\{G_{\Delta_+}(z,z)-
\;G_{\Delta_-}(z,z)\}~.\label{Schwinger}\eeq

 One might be tempted to factorize away the volume (usual
procedure) and work further only with the effective potential.
However, the perfect matching with the boundary computation will
require keeping track of the volume as well. In the spirit of the
IR-UV connection we now use dimensional regularization to control
both the IR divergencies in the bulk as well as the UV divergencies
on the boundary.

\subsection{Dimensionally regularized volume}

Starting from the usual representation of $\mathbb{H}^{d+1}$ in
terms of a unit ball with metric\\ $ds^2=4(1-x^2)^{-2}dx^2$ one
gets, after the substitution $r=(1-\vert x\vert)/(1+\vert x\vert)$,
the metric
\beq G= r^{-2}[(1-r^2)^2 g_0 + dr^2]\,,\eeq  with $4 g_0$
being the usual round metric on $\mathbb{S}^d$. Then \beq
\left(\frac{\mbox{det}\, G}{\mbox{det}\,
g_0}\right)^{^\frac{1}{2}}=r^{-1-d}\;(1-r^2)^d,\eeq and the volume
 is then given by\beq \int d{\it
vol}_{\mathbb{H}^{d+1}}\,=\,2^{-d}\,{\it vol}_{\mathbb{S}^d}
\int_0^1dr\,r^{-d-1}\;(1-r^2)^d.\eeq
 Up to this point, we have just followed \cite{Gra99} to compute the
volume. From here on, there are two standard ways to proceed in the
mathematical literature, namely Hadamard or Riesz regularization (
see, e.g.~\cite{Alb05}). We will use none of them, although our
choice of dimensional regularization is closer to Riesz's scheme.
This IR-divergent volume will now be controlled with DR~\footnote{
Some virtues of dimensional regularization in relation with
holographic anomalies were already apparent in~\cite{ISTY99,ST00}.}:
set $d\rightarrow D=d-\epsilon$ and perform the integral to get,
after some manipulations, \beq {\it
vol}_{\;\mathbb{H}^{D+1}}\,=\,\pi^{\frac{D}{2}}\;
\Gamma(-\frac{D}{2})~.\label{regvol}\eeq Let us now send $\epsilon$
to zero: \beq {\it
vol}_{\;\mathbb{H}^{D+1}}\,=\,\frac{\mathcal{L}_{d+1}}{\epsilon} +
\mathcal{V}_{d+1}+ o(1)~.\eeq For $even$ $d$ we find the
``integrated conformal anomaly'' (integral of Branson's Q-curvature,
a generalization of the scalar curvature, see e.g.~\cite{Gra99}) and
renormalized volume given by\beq
\mathcal{L}_{d+1}=(-1)^{\frac{d}{2}}\,\frac{2\pi^{\frac{d}{2}}}
{\Gamma(\frac{d+2}{2})}\;,\eeq

\beq \mathcal{V}_{d+1}\,=
\,\frac{1}{2}\,\mathcal{L}_{d+1}\cdot\left[\psi(1+\frac{d}{2})-\mbox{log}\,\pi\right]
\label{regvol-even}~.\eeq For $odd$ $d$ in turn, $
\mathcal{L}_{d+1}$ vanishes and the renormalized volume is given by
\beq
\mathcal{V}_{d+1}=(-1)^{\frac{d+1}{2}}\,\frac{\pi^\frac{d+2}{2}}
{\Gamma(\frac{d+2}{2})}~.\eeq

The {\em conformal invariants} $\mathcal{L}_{d+1}$ for $d=even$ and
$\mathcal{V}_{d+1}$  for $d=odd$ coincide with those obtained by
Hadamard regularization~\cite{Gra99}. For even $d$ in turn, the
regularized volume fails to be conformal invariant and its
integrated infinitesimal variation under a change of representative
metric on the boundary is precisely given by
$\mathcal{L}$~\cite{Gra99}. In all, the presence of the pole term is
indicative of an anomaly under conformal transformation of the
boundary metric. As usual in DR, the pole corresponds to the
logarithmic divergence in a cutoff regularization. In the pioneering
work of Henningson and Skenderis~\cite{HS98}, an IR cutoff is used
and the anomaly turns out to be the coefficient of the
$\mbox{log}\,\epsilon$ after the radial integration is performed.

\subsection{Dimensionally regularized one-loop effective
potential}

As effective potential we understand the integrand in
(\ref{Schwinger}), i.e. \beq\label{a} \frac{\partial}{\partial
m^2}\;\left ( V^+_{d+1}- V^-_{d+1} \right
)=\frac{1}{2}\;\{G_{\Delta_+}(z,z)- \;G_{\Delta_-}(z,z)\},\eeq where
the propagator at coincident points, understood as analytically
continued~\cite{KB98} from $D=d-\epsilon$, is given by  (
$m^2=\Delta (\Delta -d),~~\Delta _{\pm}=d/2 \pm\nu$) \beq
G_{\Delta}(z,z)=\frac{\Gamma(\Delta)}{2^{1+\Delta}\pi^{\frac{D}{2}}
\Gamma(1+\Delta-\frac{D}{2})}\;F(\frac{\Delta}{2},\frac{1+\Delta}{2};
1+\Delta-\frac{D}{2};1).\eeq Using now Gau{\ss}'s  formula for the
hypergeometric with unit argument and Legendre duplication formula
for the gamma function (appendix~\ref{formulae}), the dimensionally
regularized version of (\ref{a}) can be written as  \beq
\label{W_D}\frac{\partial}{\partial m^2}\;\left ( V^{+}_{D+1}-
V^{-}_{D+1}\right ) =\frac{1}{2^{D+2} \,\pi^{\frac{D+1}{2}}}\;
\Gamma(\frac{1-D}{2})\; \left [
\frac{\Gamma(\nu+\frac{D}{2})}{\Gamma(1+\nu-\frac{D}{2})}-
(\nu\rightarrow -\nu)\right ]~.\eeq Letting now $\epsilon\rightarrow
0$, the limit is trivial to take when $d=even$ since all terms are
finite; for $d=odd$ however, care must be taken to cancel the pole
of the gamma function with the zero coming from the expression in
square brackets in that case. This is in agreement with general
results of QFT in curved space; using heat kernel and dimensional
regularization one can show that in odd-dimensional spacetimes the
dimensionally regularized effective potential is finite, whereas in
even dimensions the UV singularities  show up as a pole at the
physical dimension (see, e.g.,~\cite{DeW03}), which cancel in the
difference taken above.
 Ultimately, one gets a finite result valid
for both even and odd $d$

\beq \label{Plancherel} \frac{\partial}{\partial m^2}\;\left (
V^{+}_{d+1}-V^{-}_{d+1}\right ) \,
=\,\frac{1}{2\nu}\,\frac{1}{2^d\,\pi^{\frac{d}{2}}}\;
\frac{(\nu)_{\frac{d}{2}}\,(-\nu)_{\frac{d}{2}}}{(\frac{1}{2})_{\frac{d}{2}}}
\,\equiv\mathcal{A}_d(\nu).\eeq We used the last equation also to
introduce an abbreviation $\mathcal{A}_d(\nu)$ for later
convenience. That this formula comprises both even and odd $d$ can
be better appreciated in the derivation given in
appendix~\ref{convolution}. Written in the form (\ref{Plancherel}),
this result coincides with that of Gubser and Mitra (eq. 24
in~\cite{GM03}) but now valid for $d$ odd as well. We have to keep
in mind to undo the derivative at the end. Interestingly, for the
corresponding integral the integrand is essentially the {\em
Plancherel measure}~\footnote{Presumably, the easiest way to see
this is via the spectral representation in terms of spherical
functions (see e.g.~\cite{Camp91}), it pick ups the residue at
$i\,\nu$. But the construction is valid only for the $\Delta_+$
propagator, $\Delta_-$ is only reached at the end by suitable
continuation. These details will be presented elsewhere.} for the
hyperbolic space at imaginary argument $i\,\nu$ (see
appendix~\ref{convolution}).

\subsection{Dimensionally regularized one-loop
 effective action}

The product of the regularized volume (\ref{regvol}) and the
regularized one-loop potential (\ref{W_D}) yields the dimensionally
regularized one-loop effective action

\bea \frac{\partial}{\partial m^2}\;\left
(S^{+}_{D+1}-S^{-}_{D+1}\right ) &=& \frac{1}{2}\,
\Gamma(-D)\left[\frac{\Gamma(\nu+\frac{D}{2})}{\Gamma(1+\nu-\frac{D}{2})}-
(\nu\rightarrow-\nu)\right]\nonumber\\
&= &\frac{\sin\pi\nu}{2\sin\pi D/2}~\frac{\Gamma(\frac{D}{2}+\nu
)\Gamma(\frac{D}{2}-\nu)}{\Gamma(1+D)}~.\label{Bingo}\eea The poles
of $\Gamma(-D)$ are deceiving. For $D\rightarrow odd$, the pole is
canceled against a zero from the square bracket. Only at
$D\rightarrow even$ there is a pole.

The claim now is that this full result can be recovered from the
dual boundary theory computation if we use the same regularization
procedure, namely dimensional regularization.
\section{The boundary computation: deformed partition function}

Putting the CFT on the sphere $\mathbb{S}^d$ with radius $R$, the
kernel $\Xi$ becomes \cite{GK03}

\beq \Xi =\delta^d(x,x')+\frac{f}{s^{2\Delta_-}(x,x')}\,,\eeq where
$s$ is the chordal distance on the sphere. The quotient of the
partition functions in the $\alpha-$ and $\beta-$theory is then
given by \beq \label{KK} W^+_d-W^-_d \,\equiv\,-\mbox{log}
\frac{Z_{\beta}}{Z_{\alpha}}=\frac{1}{2}\, \lim_{f\rightarrow\infty}
\mbox{log det}\;\Xi =\lim_{f\rightarrow\infty}\frac{1}{2}
~\sum_{l=0}^{\infty}\mbox{deg}(d,l) \;\mbox{log}(1+f\,g_l)~,\eeq
where \beq
g_l=\pi^{\frac{d}{2}}2^{2\nu}\frac{\Gamma(\nu)}{\Gamma(\frac{d}{2}-\nu)}
\frac{\Gamma(l+\frac{d}{2}-\nu)} {\Gamma(l+\frac{d}{2}+\nu)}~
R^{2\nu}~,\eeq
\beq\mbox{deg}(d,l)=(2l+d-1)\frac{(l+d-2)!}{l!\,(d-1)!}\equiv
\frac{2l+d-1}{d-1}\frac{(d-1)_l}{l!}~.\eeq Here $g_l$ is the
coefficient~\footnote{There is a missing factor of $2^{-\Delta}$ in
eqs. (19) and (24) of~\cite{GK03}. It can be traced back to the
chordal distance in term of the azimuthal angle
$s^2=2(1-cos\theta)$.} of the expansion of $s^{-2\Delta}(x,x')$ in
spherical harmonic and $\mbox{deg}(d,l)$ counts the degeneracies.
For large $l$ one finds (our interest concerns $0\leq\nu<1$, see
(\ref{delta})) \beq \mbox{deg}(D,l)\propto l^{D-1}~,~~~~~g_l\propto
l^{-2\nu}\label{asymp}~,\eeq implying convergence of the sum in
(\ref{KK}) for $D<2\nu$. To define (\ref{KK}) for the physically
interesting positive integers $d$ we favor dimensional
regularization and use analytical continuation from the save region
$D<0$. There, in addition, the limit $f\rightarrow\infty$ can be
taken under the sum.

For this limit an amusing property of the sum of the degeneracies
$\mbox{deg}(D,l)$ alone turns out to be crucial. After short
manipulations it can be casted into the binomial expansion of
$(1-1)^{-D}$ (see (\ref{zero})) which is zero for negative $D$, i.e.
\beq \sum_{l=0}^{\infty}\mbox{deg}(D,l)~=~0~. \label{degnull}\eeq As
a consequence all factors in $g_l$ not depending on $l$ have no
influence on the limit $f\rightarrow\infty$ and we arrive at
\beq
\label{sum} W^+_D-W^-_D
\,=\,-\frac{1}{2}\sum_{l=0}^{\infty}\mbox{deg}(D,l)\;\mbox{log}
\frac{\Gamma(l+\frac{D}{2}+\nu)}{\Gamma(l+\frac{D}{2}-\nu)} ~.\eeq
Here we want to stress that this is our full answer, whereas it is
just a piece in~\cite{GK03} where zeta-function regularization was
preferred. Although the zeta-function regularization of the sum of
the degeneracies alone vanish in odd dimensions, in even dimension
it is certainly nonzero.

To make contact with the mass derivative of the effective action of
the previous section we take the derivative
$\frac{\partial}{\partial m^2}=
\frac{1}{2\nu}\frac{\partial}{\partial \nu}$ \beq
\frac{1}{2\nu}\frac{\partial}{\partial \nu}\;\left
(W^+_D-W^-_D\right
)\,=-\frac{1}{4\nu}\;\sum_{l=0}^{\infty}\mbox{deg}(D,l)\;\left (
\psi(l+\frac{D}{2}+\nu)+\psi(l+\frac{D}{2}-\nu) \right ) ~.\eeq The
task is now to compute the sum. For this we want to exploit
Gau{\ss}'s  integral representation for $\psi(z)$
(\ref{psigauss}). However, since it requires $z>0$ we first keep
untouched the $l=0$ term and get for $2\nu -2<D<0$ \bea
\frac{1}{2\nu}\frac{\partial}{\partial \nu}\;\left(
W^+_D-W^-_D\right)&=&-\frac{1}{4\nu}
\left(\psi(\frac{D}{2}+\nu)+\psi(\frac{D}{2}-\nu) \right)\\
&& -\frac{1}{4\nu}\sum _{l=1}^{\infty}\mbox{deg}(D,l)\int
_0^{\infty}dt \left(
2~\frac{e^{-t}}{t}~-~\frac{e^{-t(l+d/2)}}{1-e^{-t}}(e^{-t\nu}+e^{t\nu})\right)~.
\nonumber \eea Now the sum of the $l$ independent term under the
integral can be performed with (\ref{degnull}). The other sums via
(\ref{zero}) can be reduced to $\sum
_{l=1}^{\infty}\frac{(D-1)_l}{l!}e^{-tl}=(1-e^{-t})^{1-D}-1~.$ Then
with $\psi (z)=\psi (1+z)-1/z$ and the Gau{\ss} representation for
$\psi (D/2+1\pm\nu)$ we arrive at \bea
\frac{1}{2\nu}\frac{\partial}{\partial
\nu}\;\left (W^+_D-W^-_D\right ) &=&\frac{D}{\nu (D-2\nu)(D+2\nu)}~+\\
&+& \frac{1}{4\nu} \int
_0^1du~u^{\frac{D}{2}-1}(u^{\nu}+u^{-\nu})\Big
((1-u)^{-D-1}(1+u)-1\Big ) ~.\nonumber \eea

In identifying the remaining integral as a sum of Beta functions
a bit of caution is necessary since we are still confined to the
convergence region $2\nu -2<D<0$. But using both the standard
representation and the subtracted version (\ref{betasub}) we finally
get

\beq \frac{1}{2\nu}\frac{\partial}{\partial \nu}\;\left
(W^+_D-W^-_D\right )\,=\, \frac{1}{2}\,
\Gamma(-D)\left[\frac{\Gamma(\nu+\frac{D}{2})}{\Gamma(1+\nu-\frac{D}{2})}-
(\nu\rightarrow-\nu)\right]~. \eeq

This has been derived by allowed manipulations of convergent sums and
integrals in the region $2\nu -2<D<0$. From there we analytically continue
and a comparison with (\ref{Bingo}) gives now for all $D$

\beq \frac{1}{2\nu}\frac{\partial}{\partial \nu}\; \left
(W^+_D-W^-_D\right )\,=\,\frac{\partial}{\partial m^2}\;\left
(S^{+}_{D+1}-S^{-}_{D+1}\right )~.\label{Patá a la lata}\eeq

 Let us just mention that one can, in principle, choose Hadamard
regularization, i.e. subtract as many terms of the Taylor expansion
in $\nu$ of the sum (\ref{sum}) as necessary to guarantee
convergence. This results in the renormalized bulk result plus a
polynomial in $\nu$ of degree $d$. This polynomial is just an
artifact of the regularization scheme and is of no physical meaning.
The question whether in this framework there is a subtraction scheme
on the boundary that exactly reproduces the bulk result seems to
find an answer in a generalization of Weierstrass formula for the
multigamma functions~\cite{Vig79}. Surprisingly, the effective
potential in AdS can be written in terms the multigamma
functions~\cite{KB98,DD06}. We refrain from pursuing this {\em
Weierstrass regularization} here and stick to DR for simplicity.
\section{Back to the physical dimensions}

Let us now send $\epsilon\rightarrow 0$ in the dimensionally
regularized partition functions (eqs. \ref{Bingo} and \ref{Patá a la
lata}) and see what happens in odd and even dimensions.

\subsection{d=odd: renormalized partition functions}

Let us assume a minimal subtraction scheme to renormalize and
establish the holographic interpretation of the boundary result. In
this case we have  \beq \frac{1}{2\nu}\frac{\partial}{\partial
\nu}\;\left (W^+_D-W^-_D\right )\,=\,
\frac{\pi}{2\nu}\,\frac{(-1)^{\frac{d+1}{2}}}{\Gamma(1+d)}\;
(\nu)_{\frac{d}{2}}\,(-\nu)_{\frac{d}{2}} + o(1).\eeq Now, the
renormalized value is exactly the renormalized volume times the
renormalized effective potential~\footnote{This exact agreement can
be upset if different regularization/renormalization procedures were
chosen, but in any case this ambiguity would show up only as a
polynomial in $\nu$ of degree $d$ at most. This is related to the
fact that if one differentiate enough times with respect to $\nu$
(equivalently, $m^2$) the result is no longer divergent and
therefore ``reg.-scheme''-independent.}

\beq \frac{1}{2\nu}\frac{\partial}{\partial \nu}\;\left
(W^+_d-W^+_d\right )= \mathcal{V}_{d+1}\;\cdot
\mathcal{A}_d(\nu)\,=\, \frac{\partial}{\partial m^2}\;\left
(S^{+}_{d+1}-S^-_{d+1}\right )~.\label{wodd} \eeq

This completes the matching in \cite{GK03}, the finite nonzero bulk
result being indeed a finite contribution in the CFT computation
which has not been computed before. At the same time, it is not a
contribution to the conformal anomaly, this being absent for $d$ odd
as expected on general grounds.

So, holography (AdS/CFT) in this case matches the renormalized
partition functions at $O(1)$ order in $CFT_d$ and at one-loop
quantum level in $AdS_{d+1}$.

\subsection{d=even: anomaly and renormalized partition functions}

Following the same steps as above, we get for this case

\beq \frac{1}{2\nu}\frac{\partial}{\partial \nu}\;\left (
W^+_D-W^-_D\right )\,=\,
\frac{1}{\epsilon}\;\cdot\frac{1}{\nu}\,\frac{(-1)^{\frac{d}{2}}}
{\Gamma(1+d)}\;(\nu)_{\frac{d}{2}}\,(-\nu)_{\frac{d}{2}}
\,+\frac{1}{2\nu}\frac{\partial}{\partial \nu}\;\left
(W^+_d-W^-_d\right )\,+ o(1).\eeq Here we can identify the
factorized form of the term containing the pole, the contribution to
the conformal anomaly,

\beq \mbox{Res}\left[\frac{1}{2\nu}\frac{\partial}{\partial \nu}\;
\left (W^+_D-W^-_D\right )\,,\,D=d\right]\,=\,
\mathcal{L}_{d+1}\;\cdot \mathcal{A}_d(\nu) \, .\eeq Note that
according to (\ref {Plancherel}) $\mathcal{A}_d(\nu)$ is just the
derivative of the difference of the renormalized effective
potentials for the $\alpha-$ and $\beta-$CFT.

This is the proof to generic even dimension of the matching between
bulk~\cite{GM03} and boundary~\cite{GK03} computations concerning
the correction to the {\em conformal anomaly}, including the overall
coefficient.

However, there is apparently a puzzle here concerning the finite
remnant. The renormalized value
\beq\frac{1}{2\nu}\frac{\partial}{\partial \nu}\;\left
(W^+_d-W^-_d\right )=\frac{\mathcal{L}_{d+1}\;\cdot
\mathcal{A}_d(\nu)}{2}
\left\{2\,\psi(1+d)-\psi(\frac{d}{2}+\nu)-\psi(\frac{d}{2}-\nu)\right\}
\label{penultima}\eeq is certainly non-polynomial in $\nu$.

Had we computed only the renormalized effective potential, then
after subtraction of the pole we would end up with the finite result
$\mathcal{V}_{d+1}\, \cdot\mathcal{A}_d(\nu)$. But
$\mathcal{A}_d(\nu)$ is polynomial in $\nu$ and therefore it could
have been renormalized away. Yet, the CFT computation renders the
non-polynomial finite result of above that cannot be accounted for
by the renormalized effective potential, which is only polynomial in
$\nu$.

Here is that IR-UV connection enters in a crucial way, and the
non-polynomial result is obtained by the cancellation of the pole
term in the regularized volume (IR) with the $O(\epsilon)$ term in
the regularized effective potential (UV). Only in this way is the
naive factorization bypassed. In fact, one can check that the
coefficient of the non-polynomial part in the CFT computation is
precisely the $\mathcal{L}$ factor, rather than the regularized
volume $\mathcal{V}$.

That is, we have to keep track of the $O(\epsilon)$ term in the
expansion of the regularized effective
potential~(\ref{W_D}, \ref{Plancherel})

\beq \frac{\partial}{\partial m^2}\;\left ( V^{+}_{D+1}-
V^{-}_{D+1}\right )\,=\, \mathcal{A}_d(\nu)
\,+\,\epsilon\,\cdot\mathcal{B}_d(\nu)
\,+\,o(\epsilon),\label{weven} \eeq where \beq
\mathcal{B}_d(\nu)=\frac{\mathcal{A}_d(\nu)}{2}\left\{
\mbox{log}(4\pi)+\psi(\frac{1}{2}-\frac{d}{2})
-\psi(\frac{d}{2}+\nu)-\psi(\frac{d}{2}-\nu)\right\}
~,\label{next-V}\eeq is almost the non-polynomial part of above.

After using two identities for $d=even$,
$\psi(\frac{1}{2}-\frac{d}{2})=\psi(\frac{1}{2}+\frac{d}{2})$ and
then $2\psi(1+d)=2\,\log
2+\psi(\frac{1}{2}+\frac{d}{2})+\psi(1+\frac{d}{2})$ -which are the
``log-derivatives'' of Euler's reflection and Legendre duplication
formula respectively~(\ref{legendre}), one can finally write the
renormalized $CFT_d$ result (\ref{penultima}) in terms of the bulk
quantities (\ref{regvol-even}, \ref{next-V}) for $d=even$ as

\beq\label{weven1}\frac{1}{2\nu}\frac{\partial}{\partial \nu}\;
\left( W^+_d-W^-_d \right)=\mathcal{V}_{d+1}\;\cdot
\mathcal{A}_d(\nu)\;+\; \mathcal{L}_{d+1}\;\cdot
\mathcal{B}_d(\nu)\,=\, \frac{\partial}{\partial m^2}\;\left(
S^{+}_{d+1}- S^{-}_{d+1}\right)~.\eeq
\section{Miscellaneous comments}

Our main results (\ref{Patá a la lata}, \ref{wodd}, \ref{weven1})
still contain a mass derivative (equivalently, derivative with
respect to $\nu$). Integrating these equations introduces an
integration constant which cannot be fixed without further input.
Equivalently, so far we only know (see (\ref{KK})) \beq
W_d(\nu)-W_d(\nu _0)~=~-\log\left
(\frac{Z_{\beta}(\nu)}{Z_{\alpha}(\nu)}~ \frac{Z_{\alpha}(\nu
_0)}{Z_{\beta}(\nu_0)}\right )~.\label{crossratio} \eeq Beyond
dimensional regularization, in the framework of general
renormalization theory, there appear free polynomials in $\nu$
anyway. Hence fixing this constant should be part of the physically
motivated normalization conditions.

It was argued in~\cite{GM03} that both $Z_{\alpha}$ and $Z_{\beta}$
at the BF mass, i.e. $\nu =0$, should coincide; the argument given
was shown in~\cite{HR06} to apply to the vacuum energy rather than
to the effective potential and the equality was argued in a
different way, replacing the BF mass by infinity as a reference
point. We just want to point out that this procedure also has a
potential loophole, namely the integration range exceeds the window
in which the two CFTs are defined $m^2_{BF}\leq m^2<1+m^2_{BF}$.

Drawing attention by the last comment to the case $\nu =0$, another
remark is in order. Then in (\ref{KK}) the product $f\,g_l$ is ill
defined if $f$ is assumed to be $\nu-$independent, as
in~\cite{GK03,HR06}. However, if one chooses
\beq\widetilde{f}=f\;\pi^{\frac{d}{2}}\frac{\Gamma(1-\nu)}{
\nu\,\Gamma(\frac{d}{2}-\nu)}\eeq as the true $\nu-$independent
quantity, then the product $f(\nu )\,g_l(\nu)\equiv \widetilde{f}
k_l(\nu)$ is well defined at $\nu=0$. The relative factor between
$f$ and $ \widetilde{f}$ can be traced back to the conventional
normalization of the two point functions. In addition, while $f$ is
the coefficient of the relevant perturbation of the $\alpha$-CFT,
$\widetilde{f}$ appears in the parametrization  of the boundary
behavior  of the bulk theory \cite{GK03,HR06}. However,
fortunately, a switch from $f$ to $\widetilde{f}$ has no effect on
the limit $f\rightarrow\infty$ in (\ref{KK}) and the conclusions
drawn from it in the previous section. This follows from the
observation stated after eq. (\ref{degnull}): any rescaling of $f$
by a factor independent of $l$ does not affect the limit.

The issue of the integration constant discussed above leading to
(\ref{crossratio}) has still another aspect concerning the treatment
of (\ref{KK}). Starting from the formal expression for\\
$W_d(\nu)-W_d(\nu_0)$ on the r.h.s. we would get $\log
(\frac{1+fg_l(\nu)}{1+fg_l(\nu _0)})$ instead of $\log
(1+fg_l(\nu))$. Now the limit $f\rightarrow\infty$ is well defined
for each $l$. The definition of the regularized sum over $l$ could
then be done directly with the summands referring  to the
difference of the two limiting conformal theories. Finally,
differentiation with respect to $\nu $ would reproduce all our
results of section 3.

Recently, in ref. \cite{HR06} a ``kinematic explanation'' of the
equivalence of the bulk and boundary computation has been given .
There a polar basis in $\mathbb{H}^{d+1}$ was used to compute the
bulk fluctuation determinants. After inserting the eigenvalues of
the angular Laplacian one ends up with a sum over spherical
harmonics of effective radial determinants which are now
one-dimensional. Using then a proposed generalization of the
Gel'fand-Yaglom formula~\cite{GY59}, it results in the same
expansion as obtained on the boundary (\ref{KK}). Since, especially
in their reasoning, it should be crucial to have a well defined
limit $f\rightarrow\infty$ {\it before} the sum over $l$ is taken,
we would prefer to consider the cross-ratios
\beq\frac{\mbox{det}_{\widetilde{f}_1}(-\Delta_{rad}+\mathcal{V}_{
eff}(\nu))}
{\mbox{det}_{\widetilde{f}_1}(-\Delta_{rad}+\mathcal{V}_{eff}(\nu_0))}
\;~\cdot~\;\frac{\mbox{det}_{\widetilde{f}_2}(-\Delta_{rad}+\mathcal{V}_{
eff}(\nu_0))}
{\mbox{det}_{\widetilde{f}_2}(-\Delta_{rad}+\mathcal{V}_{eff}(\nu))}
\label{double} \eeq instead of the single ratios obtained by
dropping the $\nu_0$ determinants. Besides giving a well defined
limit for $\widetilde{f}_1\rightarrow \infty
~,~~\widetilde{f}_2\rightarrow 0 $ this has the additional benefit
that no generalization of the Gel'fand-Yaglom formula beyond that in
\cite{KMc04} is needed to handle the ratio for operators with
different boundary conditions; each of the two ratios in
(\ref{double}) refer to the same boundary condition. Even though the
above recipe makes finite the quotient of the effective radial
determinants, the inclusion of the infinite tower of harmonics makes
the sum divergent. This remaining divergence is then the only source
for IR divergence in the bulk and UV ones on the boundary.
 The {\em formal}  equality calls for a more ambitious program
including generic dimension and not only the matching of the
anomalous part. There is nothing in the derivation that picks out
$d=even$ in preference to $d=odd$. What we have shown in the
previous section is that the equality can indeed be made rigorous if
interpreted in the sense of dimensional regularization.
\section{Conclusion}

The relevant double-trace deformation of a CFT and its AdS dual
picture provide a satisfactory test of the correspondence. The
regimes in which the bulk and boundary computations are legitimate
fully overlap and the mapping goes beyond the original correction to
the conformal anomaly. Rather the full change in the partition
functions is correctly reproduced on either side of the
correspondence; on the boundary being subleading $O(1)$ in the large
$N$ limit, and in the bulk being a 1-loop quantum correction to the
classical gravity action. Dimensional regularization proved to be
the simplest and most transparent way to control the divergences on
both sides, UV and IR infinities are then on equal footing, in
accord with the IR-UV connection.

The anomaly turns out to be the same computed in DR and in
zeta-function regularization, confirming its stability with respect
to changes in the regularization method used~\cite{DeW03}. They
differ, however, in the regularized value of the boundary
determinant for even dimension; this is also known to be the case in
free CFTs in curved backgrounds when computing the regularized
effective potential~\cite{DeW03}. Going back to the anomaly, we
recall that it arises in DR due to a cancellation of the pole
against a zero, in fact minus the variation of the counterterms is
the variation of the renormalized effective
action~\cite{ST00,DeW03}. The pole we already had from the volume
regularization and the zero comes from the invariance of
$\mathcal{L}$~\cite{Gra99}.

At odd $d$, the finite non-zero bulk change in the effective
potential is reproduced from a finite remnant in the boundary
computation, confirming the suspicion in~\cite{GK03}. But there is
no anomaly in this case, just a conformally  invariant
renormalized contribution to the partition functions. At even $d$,
in turn, the boundary change in the partition function is obtained
only after a subtle cancellation of the pole in regularized volume
(IR-div.) against a zero from the change in the effective potential
(UV-div.). This mixing of IR and UV effects on the same side of the
correspondence has no precedent in the leading order
computation~\cite{HS98}. In that case the bulk computation is a tree
level one, where no UV problems show up; i.e. the AdS answer is
obtained from the classical SUGRA action.

We can contemplate several extensions of the program carried out.
One can try to access to an intermediate stage of the RG flow, that
is, finite $f$. Being away from conformality, the factorization of
the volume breaks downs, the propagators at coincidence points
depend on the radial position; this makes the task of regularization
more difficult. Extensions to other bulk geometries seems, naively,
immediate in terms of the Plancherel measure, it admits a readily
generalization to symmetric spaces~\cite{Hel94}. It would be
interesting to explore whether this construction admits a
holographic interpretation. In the other direction, one can trade
the round sphere by a ``squashed'' one, conformal boundary of
Taub-Nut-$AdS$ and Taub-Bolt-$AdS$ spacetimes.

Finally, on the basis of the impressive agreement, one may wonder
whether there is a parallel computation in the mathematical
literature. If one is willing to allow for a continuation in
$\Delta_-$ so that it becomes  $d/2-k$  ($k=1,2,...,d/2$), then $\Xi
\sim\langle O_{\alpha}O_{\alpha}\rangle \sim1/s^{2\Delta_-}$ can be
thought of as the inverse of the $k$-th GJMS conformal
Laplacian~\cite{GJMS92}. These are conformally covariant
differential operators whose symbol is given the $k$-th power of the
Laplacian; for $k=1$ one has just the conformal Laplacian (Yamabe
operator), $k=2$ corresponds to the Paneitz operator, etc. For even
$d+1$, we find then an analogous result in theorem 1.4
of~\cite{Gui05} for a generalized notion of determinant of the
$k$-th GJMS conformal Laplacian. The absence of anomaly for odd $d$
is consistent with this determinant being a conformal invariant of
the conformal infinity of the even dimensional asymptotically
hyperbolic manifold. Unfortunately, ``the delicate case of $d+1$ odd
where things do not renormalize correctly'', is still to be
understood in this mathematical setting. We anticipate, by analogy
with our results, that a proper analysis in this case should unravel
a conformal anomaly which can be read off from quotient formulas
(generalized Polyakov formulas, see e.g.~\cite{Bra95}) of
determinants of GJMS operators at conformally related metrics, which
involve Q-curvature.

We expect the AdS/CFT recipe to treat double-trace deformations and
its bulk interpretation to be a way into these constructions in
conformal geometry~\footnote{ GJMS operators have already appeared
in matter contributions to the conformal anomaly~\cite{PS04}.}. The
leading large-N anomaly matching already hinted in this direction,
the relation between Q-curvature and volume renormalization emerges
there; but higher-dimensional Q-curvatures in connection with
generalized Polyakov formulas for GJMS operators had not shown up so
far.
\acknowledgments We have benefited from common seminars with the
group of Differential Geometry and Global Analysis at HU-Berlin, in
particular those given by J. Erdmenger and A. Juhl. We also thank A.
Wipf for sharing his insight on Weierstrass regularization and H. S.
Yang for discussions. This work was partially supported by DFG via
DO 447/4-1 and IMPRS for {\it Geometric Analysis, Gravitation and
String Theory}.

\begin{appendix}
\section{Convolution of bulk-to-boundary propagators}
\label{convolution}

The bulk-to-bulk propagator and bulk-to-boundary propagators in AdS
(see, e.g.,~\cite{KW99}), in Poincare coordinates

\beq \label{Poincare}
ds^2=\frac{1}{z_0^2}(dz_0^2+d\,\overrightarrow{x}^2),\eeq
 are given, respectively, by
\beq
G_{\Delta}(z,w)=C_{\Delta}\,\frac{2^{-\Delta}}{2\Delta-d}\,\xi^{\Delta}\;
F(\frac{\Delta}{2},\frac{\Delta+1}{2};\Delta-\frac{d}{2}+1;\xi^{-2})
,\eeq
 in term of the hypergeometric function, and
\beq
K_{\Delta}(z,\overrightarrow{x})=C_{\Delta}\,\left(\frac{z_0}{z_0^2+
(\overrightarrow{z}-\overrightarrow{x})^2}\right)^{\Delta} ,\eeq
with the normalization constant \beq
C_{\Delta}=\frac{\Gamma(\Delta)}{\pi^{\frac{d}{2}}\,
\Gamma(\Delta-\frac{d}{2})}. \eeq The quantity
\beq\xi=\frac{2\,z_0\,w_0}{z_0^2+w_0^2+(\overrightarrow{z}-
\overrightarrow{w})^2}\eeq is related to the geodesic distance
$d(z,w)=\mbox{log}\frac{1+\sqrt{1-\xi^2}}{\xi}$.

 There is a natural way to get precisely the difference of the
two bulk-to-bulk propagators of conjugate dimension in AdS/CFT. It
is based on the observation that the convolution along a common
boundary point of two bulk-to-boundary propagator of conjugate
dimensions, that is $\Delta$ and $d-\Delta$, results in the
difference of the corresponding two bulk-to-bulk
propagators~\cite{HR06,LMR03}. More precisely, the result is in fact
the sum \beq \int_{\mathbb{R}^d}d^d\overrightarrow{x}\;
K_{\Delta}(z,\overrightarrow{x})\;K_{\Delta}(w,\overrightarrow{x})=
(2\Delta-d)G_{\Delta}(z,w)+ [\Delta\leftrightarrow d-\Delta ]. \eeq

 The coincidence limit $w\rightarrow z$ can be taken before the convolution,
 on both bulk-to-boundary propagators, to get
 \beq
(2\Delta-d)\left\{G_{\Delta}(z,w)-G_{d-\Delta}(z,w)\right\}
=C_{\Delta}C_{d-\Delta}
\int_{\mathbb{R}^d}d^d\overrightarrow{x}\frac{z_0^d}{[z_0^2+
(\overrightarrow{z}-\overrightarrow{x})^2]^d}.\eeq The $z_0$
dependence in the integral is just illusory, the result is just
$\frac{2\pi^{\frac{d}{2}}}{2^d \Gamma(\frac{d+1}{2})}$. As noted by
Dobrev~\cite{Dob98}, the product of the two bulk-to-boundary
normalization factors $C_{\Delta}\,C_{d-\Delta}$ coincides, modulo
factors independent of $\Delta=\frac{d}{2}+x$, with the Plancherel
measure for the d+1-dimensional hyperbolic space evaluated at
imaginary argument $i\,x$. After putting all together, equation
(\ref{Plancherel}) is confirmed.

\section{Useful formulas}
\label{formulae} Here we collect some formulas that have been used
throughout the paper. They can all be found e.g. in \cite{AAR}.

\beq (Euler's\; reflection
)\quad\Gamma(z)\,\Gamma(1-z)\,=\,\frac{\pi}{\sin(\pi z)}\eeq

\beq (Pochhammer\;symbol)\quad
(z)_n\,=\,\frac{\Gamma(z+n)}{\Gamma(z)}\eeq

\beq (1+z)_{-n}\,=\,\frac{(-1)^n}{(-z)_n}\eeq

\beq (Legendre
\;duplication)\quad\Gamma(z)\,\Gamma(z+\frac{1}{2})\,=\,
2^{1-2z}\,\Gamma(\frac{1}{2})\,\Gamma(2z)\label{legendre}\eeq

\beq (Gauss's\;hypergeometric\;theorem, \,Re(c-a-b)>0)\quad
F(a,b;c;1)=\frac{(c-b)_{-a}}{(c)_{-a}}\eeq

\beq (Gauss's\;integral\;representation)\quad\psi(z)\,=\,
\int_0^{\infty}dt \,\left(\,\frac{e^{-t}}{t}-\frac{e^{-t\,z}}
{1-e^{-t}}\,\right)\label{psigauss}\eeq

\beq (Binomial\; expansion)\quad
(1-x)^a=\sum_{n=0}^{\infty}\frac{(-a)_n}{n!}\,x^n\label{zero}\eeq

\beq
B(a,b)-B(a,c)~=~\int _0^1du~(1-u)^{a-1}(u^{b-1}-u^{c-1})~,~~~ a>-1,~~~b,c>0~.
\label{betasub}\eeq

\end{appendix}


\end{document}